\documentclass[journal,comsoc]{IEEEtran}
\usepackage[ruled,lined]{algorithm2e}
\usepackage[ruled,lined]{algorithm2e}

\usepackage{amsthm}

\usepackage{xcolor, soul}
\sethlcolor{yellow}
\usepackage{xcolor}
\usepackage{balance}

\usepackage{mathrsfs}  
\usepackage{graphicx}
\usepackage{xcolor}
\usepackage{graphicx,epsf,epic,eepic,psfig,latexcad}
\renewcommand{\baselinestretch}{1}

\usepackage{subcaption}
\usepackage{amssymb,amsmath,caption}
\usepackage{cite}
\usepackage{amsmath, amssymb, amsthm}
\usepackage{graphicx}
\renewcommand{\baselinestretch}{1}
\usepackage{epstopdf}
\makeatletter
\usepackage{textcomp}
\makeatother
\usepackage{xcolor}
\begin{document}
	
	\title{Efficient Feedback Design for Unsourced Random Access with Integrated Sensing and Communication}
	 \author{

    \IEEEauthorblockN{Mohammad Javad Ahmadi, Mohammad Kazemi, and Rafael F. Schaefer}
  	  \thanks{This work of M. J. Ahmadi and R. F. Schaefer was supported by the German Research Foundation (DFG) as part of Germany’s Excellence Strategy - EXC 2050/1 - Project ID 390696704 - Cluster of Excellence “Centre for Tactile Internet with Human-in-the-Loop” (CeTI) of Technische Universität Dresden and in part by the German Federal Ministry of Education and Research (BMBF) within the national initiative on 6G Communication Systems through the research hub 6G-life under Grant 16KISK001K. 
    Mohammad Kazemi's work was supported by UKRI under the UK government’s Horizon Europe Funding Guarantee under Grant 101103430.}
    \thanks{M. J. Ahmadi and R. F. Schaefer are with the Chair of Information Theory and Machine Learning and with the Cluster of Excellence \textit{``Centre for Tactile Internet with Human-in-the-Loop (CeTI),''} Technische Universit\"at Dresden, 01062 Dresden, Germany (e-mail: \{mohammad\_javad.ahmadi,rafael.schaefer\}@tu-dresden.de). M. Kazemi is with the Department of Electrical and Electronic Engineering, Imperial College London, London SW7 2BT, UK (e-mail: mohammad.kazemi@imperial.ac.uk).}}
	\maketitle
	\begin{abstract}
We consider an unsourced random access (URA) system enhanced with a feedback mechanism that serves both communication and sensing tasks. While traditional URA systems do not incorporate feedback, we propose a novel feedback signal design that announces the decoding status of users and simultaneously enables target sensing. To design this dual-purpose feedback, we introduce a modified projected gradient descent algorithm that minimizes a weighted combination of communication and sensing errors. Simulation results show that the proposed feedback design outperforms the state-of-the-art feedback design in the URA literature. Furthermore, we illustrate the trade-off between communication and sensing capabilities, offering valuable insight into balancing these two tasks.

	\end{abstract}

	\vspace{-2.8mm}
	\section{Introduction}
    A fundamental challenge in next-generation wireless networks is enabling communication among a massive number of low-cost, unattended devices. These devices, common in machine-type communication (MTC) scenarios such as the Internet of things (IoT), typically send sporadic, short packets without the need for sustained connectivity. Traditional access protocols, which require tight coordination and identification, become inefficient under such conditions. To overcome these limitations, unsourced random access (URA) has emerged as a promising paradigm~\cite{polyanskiy2017perspective}. In URA, a large pool of users shares a common codebook and transmits messages without prior scheduling or identification. The base station (BS) aims to recover the set of transmitted messages rather than identifying which user sent which message. This unsourced and uncoordinated nature significantly reduces signaling overhead and latency, enabling URA to support millions of devices with only a small fraction active at any given time~\cite{Ozates2024Survey}. Various URA schemes have been proposed under different channel and system models, improving performance in terms of decoding reliability and scalability~\cite{Ahmadi2023RIS,Ahmadi2024RISUMA,Ahmadi2023Unsourced,Gkagkos2023FASURA,fengler2022pilot,Ozates2023Aslotted,ahmadi2021random,ahmadi2021Unsourced,Zhang2024Dictionary}.

Recent developments in sixth-generation (6G) wireless networks highlight the convergence of communication and sensing systems, leading to the emergence of the integrated sensing and communication (ISAC) paradigm. ISAC enables both functionalities to share resources such as spectrum, hardware, and signal processing chains, thus improving efficiency in terms of cost, bandwidth, and power consumption~\cite{Qi2022integrating,Nikbakht2024,Fei2024Revealing,Ahmadi2024Integrated,Ahmadipour2023,Zhang2025}. Although traditional URA schemes have focused primarily on communication, the increasing demand for multifunctional systems requires URA frameworks that can also support sensing capabilities. This shift has motivated novel approaches such as unsourced ISAC (UNISAC), which aim to jointly optimize communication and sensing performance in uncoordinated multiuser scenarios~\cite{Zhang2025,Ahmadi2024Integrated}.

One of the main shortcomings of traditional URA schemes is the lack of downlink feedback, which prevents users from knowing whether their messages have been successfully decoded and, consequently, whether they should retransmit, potentially resulting in data loss. To address this, the HashBeam scheme in~\cite{Ebert2022} introduces a feedback mechanism via downlink beamforming in a URA setup, enabling the BS to acknowledge active users without knowing their identities. The scheme uses a combination of channel coefficients and message hashes as user proxies, allowing scalable feedback transmission in multi-antenna setups.

In this letter, we investigate feedback transmission in a URA system enhanced by sensing capabilities. Specifically, we propose an efficient feedback signal design that informs URA users whether their messages have been successfully decoded, while simultaneously enabling angle estimation of certain sensing targets within a specific area around the BS. We demonstrate the effectiveness of the proposed system through simulation results, showing significant communication performance improvements over HashBeam, the state-of-the-art feedback design for URA, while maintaining reasonable sensing performance.
	\section{System Model}
\subsection{Preliminaries}
\label{secIIa}
In most MIMO URA schemes, each user's message sequence is divided into several parts, with some parts mapped to a low-dimensional codebook. For instance, in pilot-based algorithms, a small portion of the message sequence is used to select a codeword, referred to as the \emph{pilot}~\cite{Ahmadi2024RISUMA, Ahmadi2023Unsourced, Gkagkos2023FASURA, fengler2022pilot, Ozates2023Aslotted}. Similarly, in coded/coupled compressive sensing (CCS) schemes~\cite{fengler2021non, Zhang2024Unsourced, Tian2024Design}, the message sequence is divided into multiple sub-messages, each mapped to a low-dimensional codebook. In CCS schemes, the receiver can use the detected codeword corresponding to the first sub-message as the pilot.

At the receiver side, detected pilots provide an initial identification of users and enable channel estimation. It is important to note that the detection probability of pilots is significantly higher than that of the entire message sequence. For each detected pilot, the receiver attempts to decode the corresponding message sequence. As a result, two sets of detected pilots are formed: one set, $\mathcal{S}$, contains pilots whose associated message sequences are successfully decoded and satisfy the cyclic redundancy check (CRC), while the other set, $\tilde{\mathcal{S}}$, contains pilots whose associated message sequences fail the CRC check.

\subsection{Proposed System Model}
\begin{figure}
	\centering
\includegraphics[width=1\linewidth]{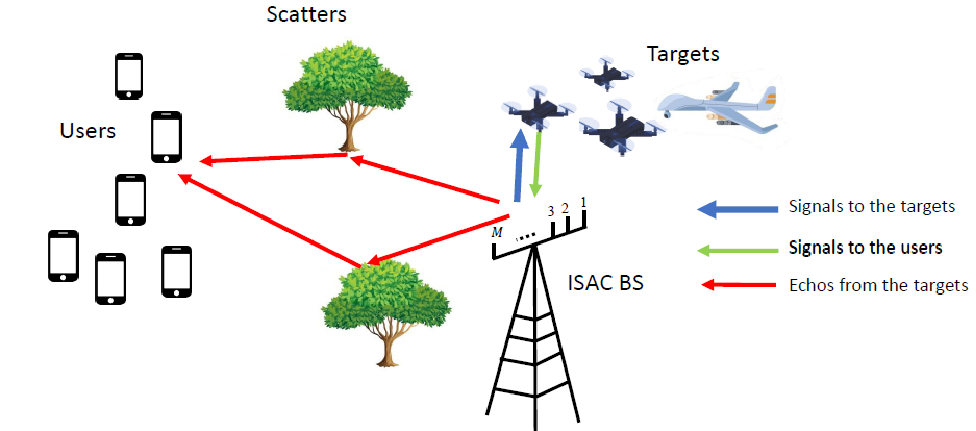}
	\caption{{\small Illustration of the ISAC model, where a downlink signal is transmitted to both users and targets.}}
	\label{Fig_configuratin}
\end{figure}

As demonstrated in Fig. \ref{Fig_configuratin}, we consider an ISAC system where a BS equipped with an \( M \)-element uniform linear array (ULA) has detected \( K = |\tilde{\mathcal{S}}| + |\mathcal{S}| \) pilots sent from the URA users, and has estimated their channel coefficient vectors, which are modeled as
 \begin{align}
     \hat{\mathbf{h}}_i=\mathbf{h}_i+\mathbf{n}_i, \label{noisyHi}
 \end{align}
where the elements of \( \mathbf{n}_i \) are randomly drawn from \( \mathcal{CN}(0, \sigma_h^2) \), and \( \mathbf{h}_i \in \mathbb{C}^M \) denotes the actual channel coefficient vector between the BS and the \( i \)th user, which is expressed as
 \begin{align}
    \mathbf{h}_i = \sum_{j=1}^N \tau_j\sqrt{\rho_0  d_i^{-\alpha_{u}}}\mathbf{a}(\theta_{ji}),
\end{align}
where \( N \) is the number of paths between each URA user and the BS, \( \tau_j \in \mathcal{CN}(0,1) \) is the amplitude of the \( j \)th path, \( \rho_0 \) denotes the path loss at unit distance, \( d_i \) is the distance between the \( i \)th user and the BS, \( \alpha_u \) is the path-loss exponent of the user-BS channels, \( \theta_{ji} \) is the angle of arrival (AOA) of the \( j \)th path of the \( i \)th user, and \( \mathbf{a}(\theta) \in \mathbb{C}^M \) is the steering vector of the ULA at the BS corresponding to the angle \( \theta \), which is of the form
\begin{align}
    \mathbf{a}(\theta)=\left[1 , e^{j\pi  \cos\theta}, e^{j2\pi  \cos \theta} , \cdots ,
e^{ j\pi (M-1)\cos \theta}\right]^T,
\end{align}
where the antenna spacing between the ULA's elements is assumed to be half of the signal wavelength. 

In the downlink phase, the BS transmits feedback to inform users about the success or failure of decoding their messages. In this way, users with detected pilots but unsuccessful message decoding have the opportunity to retransmit their signals for successful recovery, thereby reducing the overall system error. Simultaneously, the BS opportunistically utilizes the feedback signal to detect targets in its surroundings and estimate their directions by processing the reflected signals.

Considering channel reciprocity, the feedback signal received by the \( i \)th URA user, \( \mathbf{y}_i \in \mathbb{C}^{1 \times L} \), can be expressed as
\begin{align}
    \mathbf{y}_{i} = \mathbf{h}_i^T \mathbf{V}+\mathbf{z}_{i},\label{Eq_downlink}
\end{align}
where \( \mathbf{V} \in \mathbb{C}^{M \times L} \) is the transmitted signal from \(M\) BS antennas at \(L\) different channel uses with $\|\mathbf{V}\|_F^2=PL$ with \( \| \cdot \|_F \) denotes the Frobenius norm of a matrix, and \( \mathbf{z}_i \in \mathbb{C}^{1 \times L}\) denotes the additive white Gaussian noise (AWGN) vector, with each entry drawn from \( \mathcal{CN}(0, \sigma_c^2) \).

The signal echoed from \( T \) different targets and received by the BS can be written as
\begin{align}
    \mathbf{Y}_e=\sum_{t=1}^{T}\sqrt{\rho_0  d_t^{-2\alpha_t}}\mathbf{a}(\theta_t)\mathbf{a}^T(\theta_t) \mathbf{V}+\mathbf{Z}_e,\label{echoSignal}
\end{align}
where \( \theta_t \) and \( d_t \) are the angle of departure/arrival and distance between the BS and the \( t \)th target, respectively, \( \alpha_t \) is the path-loss exponent for the channels between the BS and the targets, and \( \mathbf{Z}_e \) denotes the AWGN vector, with each entry drawn from \( \mathcal{CN}(0, \sigma_e^2) \). In this equation, we assume that the distance between the BS and the targets is much smaller than that between the BS and other phenomena (e.g., users), allowing us to focus only on AWGN and targets.

The ultimate aim is to efficiently design the feedback signal, through which users are informed about the success of their transmission (communication task), and the directions of targets in the desired area are estimated (sensing task).
\section{Proposed Scheme}

\subsection{Error Analysis in Sensing and Communication Phases}
As discussed, after each uplink transmission in URA, the BS outputs \( K \) different pilots, a subset of which are successfully decoded, while others are not. For each detected pilot, whether successfully decoded or not, the receiver estimates its corresponding channel coefficient vector \(\hat{\mathbf{h}}_i\) and maps \(B_p\) bits associated with the pilot to a hash codebook \(\mathbf{P} \in \left\{ -\frac{1}{\sqrt{L}}, +\frac{1}{\sqrt{L}} \right\}^{2^{B_p} \times L}\) to generate a length-\(L\) hash vector \(\mathbf{p}_i\). In the case of successful pilot detection, this hash \(\mathbf{p}_i\) is known to both the BS and the \(i\)th user. Let \(s_i = 1\) denote successful decoding of the \(i\)th user's message and \(s_i = -1\) denote unsuccessful decoding. The \(i\)th user estimates \(s_i\) using the following decision strategy:
\begin{align}
 \hat{s}_i=
  \begin{cases}
     +1&  \mathrm{Re}\left(\mathbf{y}_i \mathbf{p}_i^H\right)\geq 0 \\
     -1& \mathrm{Re}\left(\mathbf{y}_i \mathbf{p}_i^H\right)<0 
  \end{cases}.
  \label{decision}
\end{align}
From \eqref{noisyHi} and \eqref{Eq_downlink}, we have
\begin{align}
\nonumber     \mathrm{Re}\left(\mathbf{y}_i \mathbf{p}_i^H\right)&= \mathrm{Re}\left(\mathbf{h}_i^T \mathbf{V}\mathbf{p}_i^H\right)+\mathrm{Re}\left(\mathbf{z}_i\mathbf{p}_i^H\right)\\
&= \mathrm{Re}\left(\hat{\mathbf{h}}_i^T \mathbf{V}\mathbf{p}_i^H\right)+z_i,\label{Eq_corr_extension}
\end{align}
where $z_i=\mathrm{Re}\left(-\mathbf{n}_i^T \mathbf{V}\mathbf{p}_i^H+\mathbf{z}_i\mathbf{p}_i^H\right)\sim \mathcal{CN}\left(0,\sigma_{ch}^2\right)$, with $\sigma_{ch}^2=0.5(PL\sigma_h^2+\sigma_c^2)$. To assess the communication performance, the probability of an incorrect decision can be expressed as
\begin{align}
e_{c}=\dfrac{1}{K}\sum_{i\in \{\mathcal{S},\tilde{\mathcal{S}}\}}\mathbb{P}\left(\hat{s}_i\neq s_i\right),\label{errorCom}
\end{align}
where $\mathcal{S}$ and $\tilde{\mathcal{S}}$ are defined in Section \ref{secIIa}. Based on \eqref{decision} and \eqref{Eq_corr_extension}, we obtain
\begin{align}
\nonumber
    \mathbb{P}\left(\hat{s}_i\neq s_i|  i\in \mathcal{S}\right)&=\mathbb{P}\left(\mathrm{Re}\left(\mathbf{y}_i \mathbf{p}_i^H\right)<0\right)\\\nonumber
    &=\mathbb{P}\left(\mathrm{Re}\left(\hat{\mathbf{h}}_i^T \mathbf{V}\mathbf{p}_i^H\right)+z_i<0\right)\\ 
    &=Q\left(\mathrm{Re}\left(\hat{\mathbf{h}}_i^T \mathbf{V}\mathbf{p}_i^H\right)/\sigma_{ch}\right),\label{prob1}
\end{align}
where $Q(.)$ denotes the standard $Q$-function. Similarly, for the users in the set $\tilde{\mathcal{S}}$, we get
\begin{align}
    \mathbb{P}\left(\hat{s}_i\neq s_i| i\in\tilde{\mathcal{S}}\right)&=Q\left(-\mathrm{Re}\left(\hat{\mathbf{h}}_i^T \mathbf{V}\mathbf{p}_i^H\right)/\sigma_{ch}\right).\label{prob2}
\end{align}
Plugging \eqref{prob1} and \eqref{prob2} into \eqref{errorCom}, the detection error of the communication part is obtained as
\begin{align}
\nonumber    e_{c}= &\sum_{i\in \tilde{\mathcal{S}}}\dfrac{1}{K}Q\left(-\mathrm{Re}\left(\hat{\mathbf{h}}_i^T \mathbf{V}\mathbf{p}_i^H\right)/\sigma_{ch}\right)\\
    &+\sum_{i\in \mathcal{S}}\dfrac{1}{K}Q\left(\mathrm{Re}\left(\hat{\mathbf{h}}_i^T \mathbf{V}\mathbf{p}_i^H\right)/\sigma_{ch}\right).\label{err_comm}
\end{align}
To evaluate the angle estimation performance during the sensing phase, we present an approximate lower bound on its root mean squared error (MSE). For the channel model in \eqref{echoSignal}, the lower bounds on the MSE of the angle estimation for different targets are given by the diagonal entries of the following matrix~\cite[Theorem 4.1]{StoicaMUSIC}.
\begin{align}
     \mathbf{C}= \dfrac{\sigma_e^2}{2}\left(\sum_{s=1}^{L} \mathrm{Re}\left(\mathbf{X}_s^H
   \mathbf{D}^Hf_p(\mathbf{B}_a)\mathbf{D}\mathbf{X}_s\right) 
   \right)^{-1}, \label{CRLB}
\end{align}
where $\mathbf{X}_s\in \mathbb{C}^{T\times T}$ is a diagonal matrix whose $t$th diagonal element is the $s$th entry of the vector $\sqrt{\rho_0  d_t^{-2\alpha_t}}\mathbf{a}^T(\theta_t) \mathbf{V}$ for $t=1,2,...,T$, columns of $\mathbf{B}_a\in \mathbb{C}^{M\times T}$ are $\mathbf{a}(\theta_t)$ corresponding to $t=1,2,...,T$, $f_p\left(\mathbf{B}\right)=\mathbf{I}_M-\mathbf{B}(\mathbf{B}^H\mathbf{B})^{-1}\mathbf{B}^H$
, and $\mathbf{D}\in \mathbb{C}^{M\times T}$ represents the element-wise partial derivative of $\mathbf{B}_a$ with respect to the angle $\theta$, i.e., $\mathbf{D}=\partial \mathbf{B}_a/\partial \theta$. Considering $f_p\left(\mathbf{B}\right)\preceq \mathbf{I}_n$ and assuming $\mathrm{diag}\left(\mathbf{D}\mathbf{D}^H\right)^{-1}\approx (\mathbf{D}\mathbf{D}^H)^{-1}$, an approximate lower bound of the angle estimation MSE for the $t$th target is obtained as
\begin{align}
    e_{s}(\mathbf{V},\theta_t)\approx \dfrac{V(\theta_t)}{\mathbf{a}^T(\theta_t) \mathbf{V}\mathbf{V}^H\mathbf{a}^*(\theta_t) },
\end{align}
where $V(\theta)=\dfrac{\sigma_e^2}{2 \rho_0  d_t^{-2\alpha_t}\pi^2 \sin^2{\theta}}$. The expected value of $e_{s}(\mathbf{V},\theta_t)$ is then calculated as
\begin{align}
    e_s=\dfrac{1}{|\mathcal{T}|}\sum_{\theta\in \mathcal{T}}e_{s}(\mathbf{V},\theta),\label{MSEAOACRLB}
\end{align}
where \(\mathcal{T}\) denotes the set of angles spanning the area of interest in which the targets are to be detected. In this equation, we assume that each element of \(\mathcal{T}\) occurs with a uniform distribution, i.e., \(\mathbb{P}(\theta) = \frac{1}{|\mathcal{T}|}\) for \(\theta \in \mathcal{T}\).
\subsection{Transmit Signal Design }
For signal design, i.e., selecting \(\mathbf{V}\), we jointly minimize the error metrics corresponding to the communication and sensing phases, namely, \(e_{c}\) in \eqref{err_comm} and \(e_{s}(\mathbf{V})\) in \eqref{MSEAOACRLB}. To determine the optimal $\mathbf{V}$, the following optimization problem must be solved:
\begin{subequations}
    \label{Problem1}
\begin{align}
    \min_{\mathbf{V}} \ \ &\mu \ e_{c}+(1-\mu)e_s\\ 
    &\mathrm{s.t.} \ \ \| \mathbf{V}\|_F^2 \leq PL,
\end{align}
\end{subequations}
where \( 0 \leq \mu \leq 1 \) controls the trade-off between communication and sensing objectives.

To make the problem more tractable, we simplify functions \(Q(x)\) and \(1/x\) in the cost function. Noting that \(Q(x)\) is a strictly decreasing nonlinear function that decreases very slowly once it reaches a small value, we approximately replace \(\min_x Q(x)\) with \(\min_x (-x)\) under the condition that \(Q(x)\) does not fall below a specified threshold. This approach is motivated by the observation that once a user's detection error (expressed as \(Q(x)\), as shown in \eqref{prob1} and \eqref{prob2}) falls below the desired threshold, further reductions become negligible and unnecessary. Consequently, the algorithm should ignore such users and instead prioritize those with higher error probabilities, who contribute more significantly to reducing the total detection error. Additionally, we approximate the \( 1/x \) term using a first-order Taylor expansion. These simplifications lead to the following approximate problem:
\begin{subequations}
    \begin{align}
\nonumber    \min_\mathbf{V}&\ \mu\left(\sum_{i\in \tilde{\mathcal{S}}}
    \mathrm{Re}\left(\hat{\mathbf{h}}_i^T \mathbf{V}\mathbf{p}_i^H\right)-\sum_{i\in \mathcal{S}}\mathrm{Re}\left(\hat{\mathbf{h}}_i^T \mathbf{V}\mathbf{p}_i^H\right)\right)\\
    &+(1-\mu)\sum_{\theta \in \mathcal{T}}V(\theta)\left( \frac{1}{y_{\theta}^{o}} - \frac{y_{\theta}(\mathbf{V}) - y_{\theta}^{o}}{\left(y_{\theta}^{o}\right)^2} \right),\label{problem14}\\\nonumber
    \mathrm{s.t.}\\
    &Q\left(-\mathrm{Re}\left(\hat{\mathbf{h}}_i^T \mathbf{V}\mathbf{p}_i^H\right)/\sigma_{ch}\right)>q_o, \ \ i\in\tilde{\mathcal{S}},\label{eq13b}\\
    &Q\left(\mathrm{Re}\left(\hat{\mathbf{h}}_i^T \mathbf{V}\mathbf{p}_i^H\right)/\sigma_{ch}\right)>q_o, \ \ i\in\mathcal{S},\label{eq13c}\\
    &\left\|\mathbf{V}\right\|_F^2<P  L,
\end{align}
\label{Eq14}
\end{subequations}
where \( q_o \) is the threshold for the desired error probability, \( y_{\theta}(\mathbf{V}) = \mathbf{a}^T(\theta) \mathbf{V} \mathbf{V}^H \mathbf{a}^*(\theta) \), and \( y_{\theta}^{o} \) is a feasible point for \( y_{\theta}(\mathbf{V}) \) around which the Taylor expansion is derived. Since the cost function of this problem is convex, and motivated by the projected gradient approach, we propose the following iterative algorithm.
\begin{subequations}
    \begin{align}
\mathrm{Step 1: }\  &   \mathbf{V}_{(k+1)}=\dfrac{\mathbf{V}_{(k)}}{\|\mathbf{V}_{(k+1)}\|_F}-\eta\left( \dfrac{\mu \Delta^c_k}{\|\Delta^c_k\|_F}+\dfrac{(1-\mu)\Delta^s_k}{\|\Delta^s_k\|_F}\right)\label{step1_pgd}\\
\mathrm{Step 2: } \ & \mathbf{V}_{(k+1)}=\dfrac{\mathbf{V}_{(k+1)}}{\|\mathbf{V}_{(k+1)}\|_F}P L ,\label{step2_pgd}
\end{align}
\label{algorithm}
\end{subequations}
where \( \mathbf{V}_{(k)} \) is the estimate of the matrix \( \mathbf{V} \) in the \( k \)th iteration, \( \eta \) is the step size, and \( \Delta^c_k \) and \( \Delta^s_k \) are the gradients of the first and second terms of the cost function in \eqref{problem14} with respect to \( \mathbf{V} \) at the \( k \)th iteration, which are computed as
\begin{align}
    \Delta^c_k&=\sum_{i\in \tilde{\mathcal{A}}_k}
    \hat{\mathbf{h}}_i^*\mathbf{p}_i-\sum_{i\in \mathcal{A}_k}\hat{\mathbf{h}}_i^*\mathbf{p}_i,\label{eq17}\\
    \Delta^s_k&=\sum_{\theta \in \mathcal{T}} - \frac{\mathbf{a}^*(\theta) \mathbf{a}^T(\theta)\mathbf{V}_{(k)}}{\left(\mathbf{a}^T(\theta) \mathbf{V}_{(k)} \mathbf{V}_{(k)}^H \mathbf{a}^*(\theta)\right)^2\sin^2 \theta} , \label{eq18}   
\end{align}
where in \eqref{eq18}, we have selected \( y_{\theta}^{o} = \mathbf{a}^T(\theta) \mathbf{V}_{(k)} \mathbf{V}_{(k)}^H \mathbf{a}^*(\theta) \),
$\mathcal{A}_k = \left\{ i \in \mathcal{S} \mid Q\left({ \, \mathrm{Re}\left(\hat{\mathbf{h}}_i^T \mathbf{V}_{(k)} \mathbf{p}_i^H\right)}/{\sigma_{ch}} \right) > q_k \right\}$, and $\tilde{\mathcal{A}}_k = \left\{ i \in \tilde{\mathcal{S}} \mid Q\left(-{ \, \mathrm{Re}\left(\hat{\mathbf{h}}_i^T \mathbf{V}_{(k)} \mathbf{p}_i^H\right)}/{\sigma_{ch}} \right) > q_k \right\}$ with \( q_k \) being selected as the average of the detection errors of all the users in the $(k-1)$th iteration, i.e., 
    \begin{align}
    \nonumber
    q_k = \dfrac{1}{K} \Bigg(
    & \sum_{i \in \tilde{\mathcal{S}}} Q\left(- \, \mathrm{Re}\left(\hat{\mathbf{h}}_i^T \mathbf{V}_{(k-1)} \mathbf{p}_i^H\right)/\sigma_{ch}\right) \notag \\
    & + \sum_{i \in \mathcal{S}} Q\left( \, \mathrm{Re}\left(\hat{\mathbf{h}}_i^T \mathbf{V}_{(k-1)} \mathbf{p}_i^H\right)/\sigma_{ch}\right)
    \Bigg).\label{qk_eq}
    \end{align}
    Note that in \eqref{step1_pgd}, the update terms for communication and sensing, as well as the matrix \( \mathbf{V}_{(k)} \), are divided by their Frobenius norms to ensure a stable step size, independent of their magnitudes, and to balance the contributions of the communication and sensing cost functions. The iterative algorithm in \eqref{algorithm} is stopped after $N_{stp}$ iterations, and the signal matrix is initially set as 
\begin{align}
    \mathbf{V}_{(0)}=\sum_{i\in \tilde{\mathcal{S}}_k}
    \hat{\mathbf{h}}_i^*\mathbf{p}_i-\sum_{i\in \mathcal{S}_k}\hat{\mathbf{h}}_i^*\mathbf{p}_i.\label{initialV}
\end{align}
The proposed feedback signal design is illustrated in Algorithm \ref{algorithm1}.
\begin{algorithm}
  \caption{Feedback signal design} 
  \label{algorithm1}
\textbf{Input}: $\mathbf{p}_i$ and $\hat{\mathbf{h}}_i$ for $i\in \{\mathcal{S},\tilde{\mathcal{S}}\}$, and $\mathcal{T}$.\\
Calculate $\mathbf{V}_{(0)}$ according to \eqref{initialV}.\\
			\For( {}){$k=1,2,...,N_{stp}$}
   {
   \begin{enumerate}
   \item Calculate  $q_k$ according to \eqref{qk_eq}.
   \item Calculate $\Delta^c_k$ and $\Delta^s_k$ according to \eqref{eq17} and \eqref{eq18}.       
   \item Perform Step1 and Step2 in \eqref{step1_pgd} and \eqref{step2_pgd}.       
   \end{enumerate}
   }
 \end{algorithm}
\section{Numerical Results}
This section investigates the performance of the proposed system using Monte Carlo simulations. The following parameter values are used: the number of receive antennas is \( M = 20 \); the maximum transmit power is \( P = 13\)dBm; the reference path loss is \( \rho_0 = -30 \) dBm, and the noise levels at the location of each user and the BS are \( \sigma_c^2 = \sigma_e^2 = -100 \) dBm, and the noise variance of the channel estimation is selected as $\sigma_h^2=-\infty dBm$; the number of paths between each user and the BS is \( N = 5 \); a target is located at a distance of \( d_t = 300 \) meters from the BS, with the angle of arrival/departure selected randomly and uniformly between \( 80^o < \theta < 100^o \); the distance between each user and the BS is randomly drawn between 1000 meters and 1500 meters; the set \( \mathcal{T} \) in \eqref{Problem1} is selected by choosing 20 equally spaced values between \( 80^o \) and \( 100^o \); the path-loss exponents for the BS-target and BS-user channels are selected as \( \alpha_t = 2.2 \) and \( \alpha_u = 3 \), respectively; for the iterative algorithm, the number of iterations is \( N_{stp} = 30 \) and the step size is \( \eta = 0.1 \). To assess the performance of the algorithm, we adopt the root MSE (RMSE) of the angle estimation for the targets in the sensing phase, and the probability of incorrect decision for the users in the communication phase as
\begin{align}
\delta_s &=  \sqrt{\dfrac{1}{N_{mont}}\sum_{i=1}^{N_{mont } }|\theta_i-\hat{\theta}_i |^2},\label{Eq3_14MARCH}\\
e_c &=\dfrac{n_{f}}{K}, 
\end{align}
where \( n_{f} \) is the number of users with incorrect decisions, \( N_{mont} \) is the number of Monte Carlo iterations, \( \theta_{ij} \) is the angle of the $j$th target in the \( i \)th Monte Carlo iteration, and \( \hat{\theta}_i \) is its estimated value.

\begin{figure}[t!]
		\centering
		\includegraphics[width=.975\linewidth]{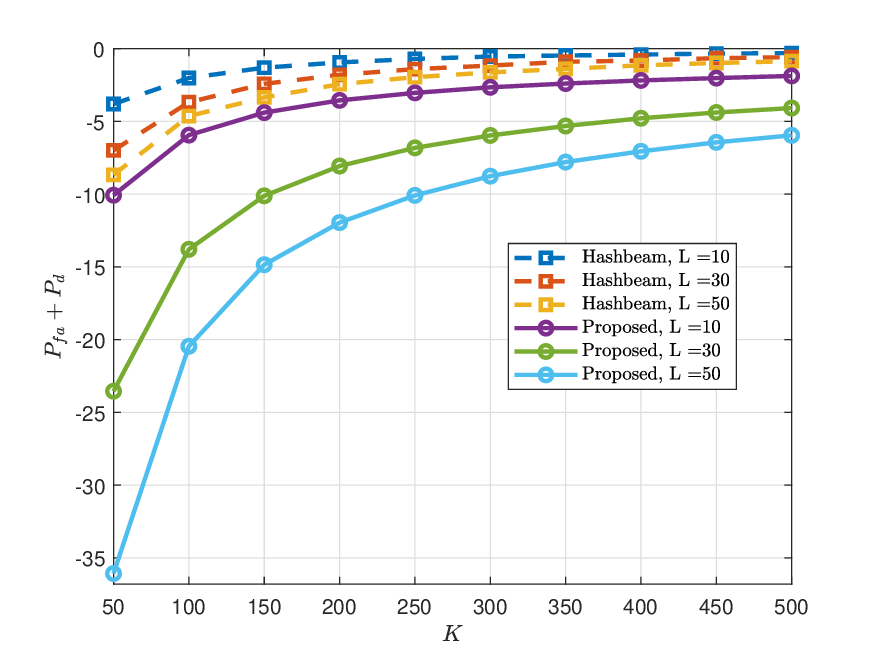}
		\caption{{\small {Performance comparison between the proposed scheme and the Hashbeam scheme in detecting successful and unsuccessful decoding, evaluated for different numbers of users \( K \) and feedback signal lengths \( L \).}
}}
		\label{Communication_only}
	\end{figure}

In Fig.~\ref{Communication_only}, the communication performance (case of \( \mu = 1 \)) of the proposed algorithm is compared with the HashBeam scheme in \cite{Ebert2022} for different values of \( L \) and \( K \), where \( |\mathcal{S}| = \lfloor 0.9 K \rfloor \) and \( |\tilde{\mathcal{S}}| = K - |\mathcal{S}| \). It is evident from this figure that the proposed signaling model results in a significantly lower detection error compared to that of HashBeam. The superior performance can be attributed to several factors, including: (1) defining a problem specifically aligned with the system’s goal---namely, minimizing \( e_{c} \)---as opposed to the heuristic linear minimum mean squared error approach proposed by HashBeam, and (2) incorporating information from both successfully and unsuccessfully decoded users in the feedback signal, unlike HashBeam, where the feedback signal was designed using only information from successfully decoded users.

\begin{figure}[t!]
		\centering
		\includegraphics[width=.975\linewidth]{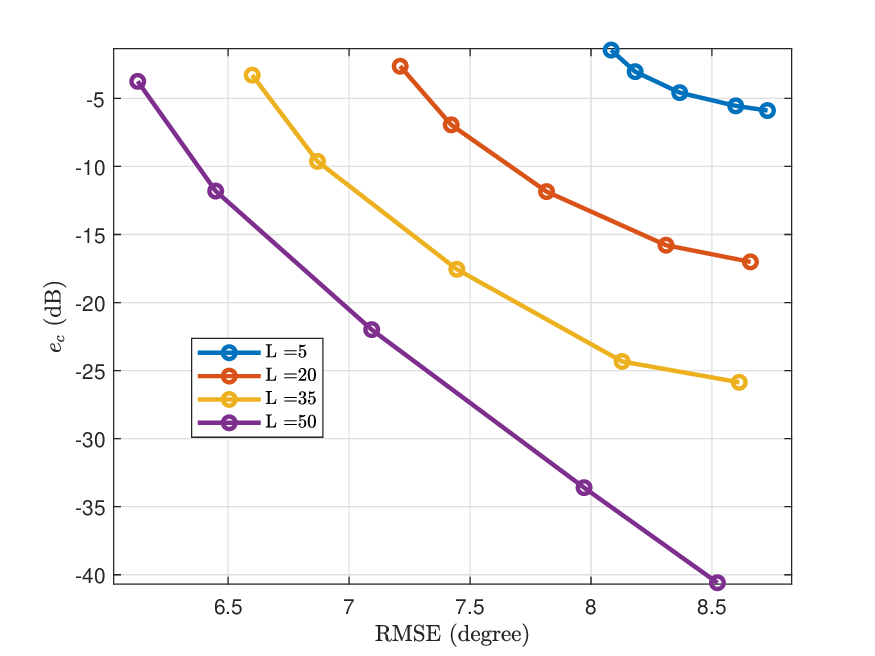}
		\caption{{\small Pareto frontier showing communication performance versus sensing performance.}}
		\label{communicaiton_sensing}
	\end{figure}

To illustrate the sensing-communication trade-off, we present the Pareto frontier between sensing and communication performance in Fig.~\ref{communicaiton_sensing}. Communication performance is measured using \( e_c \), while sensing performance is quantified by the RMSE. To generate this figure, we select five different values for \( \mu \), namely \( \mu = \{0, 0.25, 0.5, 0.75, 1\} \), set \( |\mathcal{S}| = 45 \), \( |\tilde{\mathcal{S}}| = 5 \), and vary the length of the feedback signal \( L \). From this figure, it is evident that increasing \( \mu \) improves communication performance but deteriorates sensing performance, and vice versa. Furthermore, a significant enhancement in both performances is achieved by increasing the parameter \( L \).

\section{Conclusion}
In this paper, we have presented a novel approach to feedback signal design in unsourced random access systems, where the feedback not only ensures reliable communication by informing users of their decoding success or failure, but also enables the base station to perform sensing tasks, specifically angle estimation of targets. The proposed feedback scheme efficiently minimizes both communication and sensing errors. It significantly outperforms state-of-the-art methods in communication performance while providing a flexible trade-off between communication and sensing tasks.
\balance

\end{document}